\documentclass[aps,prl,preprint,superscriptaddress]{revtex4-1}

\usepackage[swedish,english]{babel}
\usepackage[utf8]{inputenc} 
\usepackage{amsmath}
\usepackage{siunitx}
\usepackage{graphicx}

\usepackage{filecontents}

\begin{document}


\title{Stabilizing a high-pressure phase in InSb at ambient conditions with a laser-driven pressure pulse}


\author{A. Jarnac}
\affiliation{MAX IV Laboratory, Lund University, P.O. Box 118, SE-221 00 Lund, Sweden}
\affiliation{Department of Physics, Lund University, P.O. Box 118, SE-221 00 Lund, Sweden}

\author{Xiaocui Wang} 
\affiliation{Department of Physics, Lund University, P.O. Box 118, SE-221 00 Lund, Sweden}

\author{\r{A}. U. J Bengtsson} 
\affiliation{Department of Physics, Lund University, P.O. Box 118, SE-221 00 Lund, Sweden}

\author{M. Burza} 
\affiliation{MAX IV Laboratory, Lund University, P.O. Box 118, SE-221 00 Lund, Sweden}

\author{J. C. Ekstr\"om} 
\affiliation{Department of Physics, Lund University, P.O. Box 118, SE-221 00 Lund, Sweden}

\author{H. Enquist} 
\affiliation{MAX IV Laboratory, Lund University, P.O. Box 118, SE-221 00 Lund, Sweden}

\author{A. Jurgilaitis} 
\affiliation{MAX IV Laboratory, Lund University, P.O. Box 118, SE-221 00 Lund, Sweden}

\author{N. Kretzschmar}
\affiliation{ESRF The European Synchrotron, 71 Avenue des Martyrs, 38000 Grenoble, France}

\author{A. I. H. Persson}
\affiliation{Department of Physics, Lund University, P.O. Box 118, SE-221 00 Lund, Sweden}

\author{C. M. Tu}
\affiliation{Department of Physics, Lund University, P.O. Box 118, SE-221 00 Lund, Sweden}

\author{M. Wulff}
\affiliation{ESRF The European Synchrotron, 71 Avenue des Martyrs, 38000 Grenoble, France}

\author{F. Dorchies}
\affiliation{Univ. Bordeaux, CNRS, CEA, CELIA (Centre Lasers Intenses et Applications), UMR 5107, 33400 Talence, France}

\author{J. Larsson}
\affiliation{MAX IV Laboratory, Lund University, P.O. Box 118, SE-221 00 Lund, Sweden}
\affiliation{Department of Physics, Lund University, P.O. Box 118, SE-221 00 Lund, Sweden}
\email[]{jorgen.larsson@maxiv.lu.se}


\date{\today}

\begin{abstract}
In this letter, we describe the stabilization of indium antimonide (InSb) in the high-pressure orthorhombic phase (InSb-III) at ambient conditions. Until now, InSb-III has only been observed above 9 GPa, or at around 3 GPa as a metastable structure during the phase transition from cubic zinc blende (InSb-I) to orthorhombic InSb-IV. The crystalline phase transition from InSb-I to InSb-III was driven by an ultrashort, laser-generated, non-hydrostatic pressure pulse. The transition occurred in preferred orientations locked to the initial orientation of the InSb-I crystal, breaking the symmetry of the InSb-I cubic cell to form the InSb-III orthorhombic cell.
\end{abstract}


\maketitle

\begin{figure}
\includegraphics{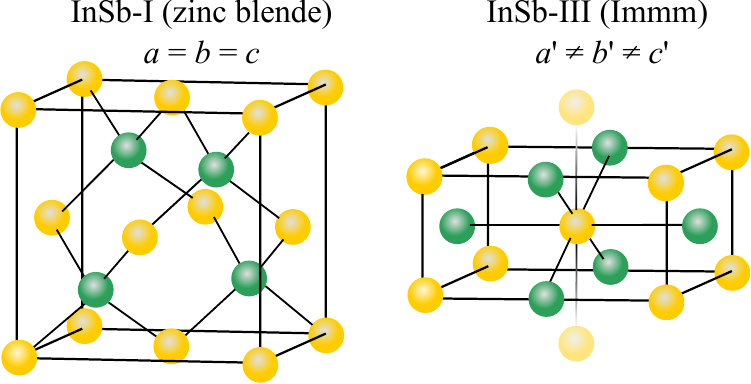}
\caption{\label{fig1}Cubic unit cell of InSb-I with a zinc blende structure and the orthorhombic unit cell of InSb-III with the space group Immm.}
\end{figure}

Controlling the structure of solid matter has long been a topic of considerable interest in materials science. High-pressure phases of III-V semiconductors are known to exhibit properties that differ significantly from those in the stable low-pressure phase, for example, metal-like properties \cite{ZhangPRB87} and superconductivity \cite{ChangPRL85}. Stabilizing the materials in these phases at ambient conditions represents a significant step forward, not only in fundamental research, but also for technological applications. Indeed, the high conductance of the high-pressure phases may pave the way for the development of new functional materials. \\
InSb has been extensively studied under high pressures \cite{MujicaRMP03}. Above 9 GPa, the thermodynamically stable phase is the orthorhombic phase InSb-III (Fig. \ref{fig1}). InSb-III has also been identified as an intermediate (metastable) phase at moderate pressures ($ \sim $ 3 GPa), during the phase transition from the cubic phase, InSb-I (Fig. \ref{fig1}) to the orthorhombic phase, InSb-IV. InSb-III has been stabilized in a fast diamond anvil cell, in which it was possible to ramp up the pressure from 0 to 3 GPa in 30 seconds, as long the pressure was maintained \cite{McWhanJCP66,NelmesPRB93,NelmesPRL96}. However, it has been found that InSb-III always relaxes back to InSb-I when the pressure returns to ambient pressure \cite{McWhanJCP66,NelmesPRB93,NelmesPRL96}. \\
Previously reported phase transitions were obtained in diamond anvil cells, where the pressure can be uniformly applied in all three dimensions (hydrostatic pressure) for periods of minutes or hours. However, it has been concluded from pseudopotential calculations that InSb-III can be stabilized at moderate pressures by applying non-hydrostatic pressure (i.e. the pressure is applied along only one dimension of the sample) \cite{KelseyJPCM00}. Moreover, molecular dynamics calculations have predicted that a phase transition will occur at 3 GPa, from the cubic to an orthorhombic phase with similar structure to InSb-III, on the picosecond time scale \cite{CostaPRB02}. We set out to investigate whether InSb-III could be stabilized using a non-hydrostatic pressure pulse with a duration similar to that of the transition time scale. \\
It is difficult to apply a non-hydrostatic pressure to a crystal on the picosecond time scale. A diamond anvil cell can be used to obtain non-hydrostatic pressure, but the pressure ramp lasts seconds or minutes \cite{McWhanJCP66}. Mechanical impact methods can be used to obtain non-hydrostatic pressure increase on a shorter time scale, of microseconds to milliseconds \cite{PizaniJPCS14}. Even shorter time scales, nanoseconds or sub-nanosecond, can be achieved using laser shocks. These have been used to study matter under extreme conditions (MEC), revealing new parts of the phase diagram that have hitherto been inaccessible \cite{GormanPRL15}. To achieve the pressure conditions required in this study, we developed an experimental system that can deliver picosecond, non-hydrostatic pressure pulses.

\begin{figure*}
\includegraphics{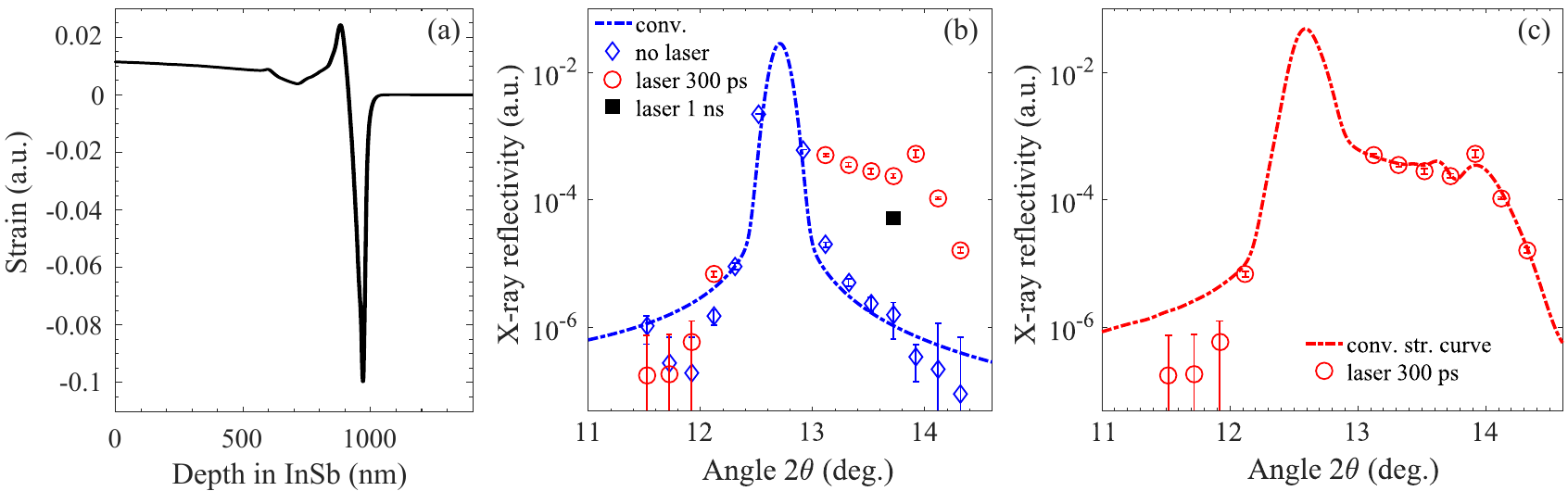}
\caption{\label{fig2} (a) Transient strain profile as a function of depth in InSb at a delay of 300 ps, calculated with the hydrodynamic code ESTHER for an absorbed laser fluence in Al of 42 mJ/cm$^{2}$ $\pm$ 3.5 mJ/cm$^{2}$. The strain pulse exhibits a compressive component (negative strain) during 25 ps, which reaches an amplitude of 10\% $\pm$ 0.5\%. (b) $\theta-2\theta$ diffraction curve from InSb before the laser pulse (blue diamonds) and after the interaction of the laser pulse with the Al film at time delays of 300 ps (red circles) and 1 ns (black square). The X-ray reflectivity was measured at least three times for all $2\theta$ angles, and the error bars show the highest and lowest intensity recorded for each angle. The dot-dashed blue curve shows the theoretical $\theta-2\theta$ curve convoluted with the angular response function of the instrument. (c) Comparison between the experimental X-ray reflectivity obtained at 300 ps (red circles) and the dynamical $\theta-2\theta$ curve calculated from the strain profile shown in (a), after convolution with a 100 ps Gaussian function (FWHM) and the angular response function of the instrument (dot-dashed red line).}
\end{figure*}

Here, we describe the stabilization of InSb in the high-pressure orthorhombic phase (InSb-III) at ambient conditions. The crystalline phase transition from InSb-I to InSb-III was triggered using a picosecond, non-hydrostatic pressure pulse. The pressure pulse was generated by partly melting a thin aluminum (Al) film deposited on a crystal of InSb using a mode-locked Ti:sapphire laser. The pressure pulse in InSb was then characterized using \textit{time-resolved} X-ray diffraction. To understand the mechanism of pressure pulse generation, the results were compared with 1D hydrodynamic simulations of the Al film. By recording the \textit{in situ} X-ray diffraction pattern from the InSb crystal after being exposed to the laser-generated pressure pulse, we found that the crystalline transition occurs in preferred orientations, breaking the symmetry of the InSb-I cubic cell to form the InSb-III orthorhombic cell. 

This study was performed at the time-resolved X-ray diffraction beamline ID09 of the European Synchrotron Radiation Facility (ESRF) \cite{WulffFD03}. A thin layer of Al (300 nm) was deposited on a (111) symmetrically cut InSb wafer. The thickness of the Al film was chosen to ensure that all the laser energy was deposited inside the film, thereby avoiding optical or thermal damage to the InSb substrate. The Al film was illuminated with 800 nm laser pulses with a duration of 1.2 ps (p-polarized). At this wavelength,  the optical reflectivity of the Al film was found to be 60\%, in agreement with the value calculated from tabulated optical parameters \cite{McPeakACSP15}. Since melting threshold fluences are difficult to determine experimentally (absorbed fluences between 5 mJ/cm$ ^{2} $ and 20 mJ/cm$ ^{2} $ are reported in ref. \cite{GuoPRL00,KandylaPRB07,SiwickScience03}), the damage threshold was used to ensure that the material was molten. Permanent damage to the Al film became apparent at an incident fluence of 120 mJ/cm$ ^{2} $, corresponding to an absorbed fluence of $ \sim $ 50 mJ/cm$ ^{2} $. To ensure melting of the Al over a sufficiently large area, the laser spot size was chosen such that the Al film was permanently damaged over an area of 1 $ \times $ 0.25 mm$ ^{2} $ (vertical (V) $ \times $ horizontal (H)). Melting of the Al film is associated with an abrupt change in density between solid Al (2709 kg/m$ ^{3} $) and liquid Al (2375 kg/m$ ^{3} $ \cite{NaschPCL95}). This change in density launches a strain pulse at the surface that propagates to the Al/InSb interface and then into the InSb crystal, where it was probed with \textit{time-resolved} X-ray diffraction. X-ray pulses with an energy of 15 keV and 100 ps duration were used, which corresponds to a Bragg angle of $ \theta_{B} $ = 6.35$^{\circ}$. The projected X-ray beam size (footprint) on the sample was 0.5 $ \times $ 0.09 mm$ ^{2} $ (V $ \times $ H). The laser pulses were electronically phase-locked to a single electron bunch in the storage ring, allowing synchronization between the laser pulse and the X-ray probe with an accuracy of 5 ps. The sample was laterally translated between successive laser pulses so that each laser pulse interacted with an undamaged area. The experiment was thus performed in single-shot mode, and the repetition rate of the X-ray pulses was reduced to 1 Hz. Additional details of the experimental setup can be found in the Supplemental Material \cite{SuppMat}.

\begin{figure*}
\includegraphics{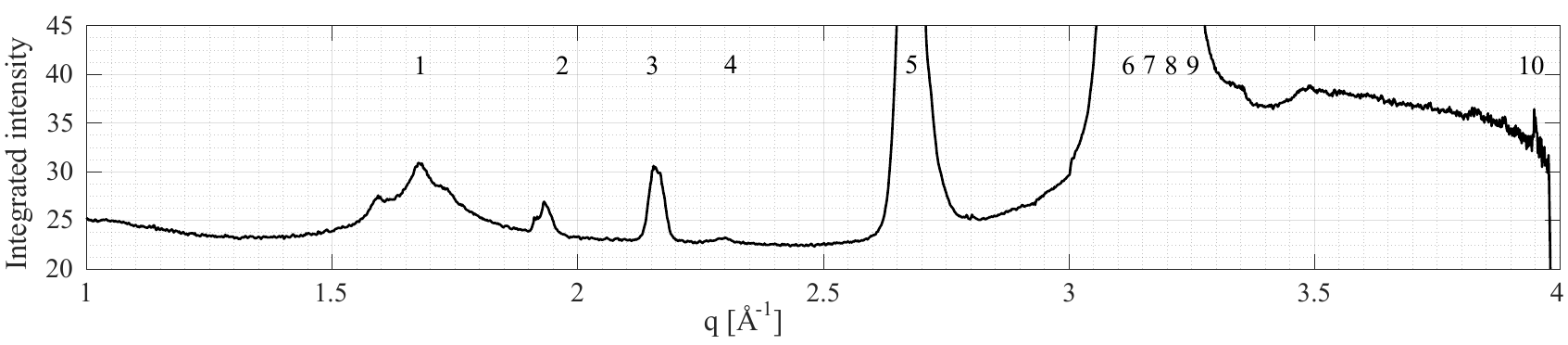}
\caption{\label{fig3} Radial integral of the \textit{in situ} diffraction pattern recorded from a laser-exposed area of the sample. The reflections are indexed with increasing \textit{q} values and are identified in Table \ref{table1}. Acquisition time: 10 seconds = 10$^{11}$ incident photons.}
\end{figure*}

\begin{table*}
\caption{\label{table1} List of reciprocal vectors (\textit{q}) of bright \textit{hkl} reflections and the lattice parameters for Al \cite{WoltersdorfSS81}, InSb-I \cite{SlutskyPR59} and orthorhombic InSb-III (previous studies) \cite{NelmesPRB93,NelmesPRL96,MezouarPSSB96}. The first column gives the InSb-III reflections observed in this work.}
\begin{ruledtabular}
\begin{tabular}{c c c c c}
 & Immm, this work & Al & zb, InSb I & Immm, InSb III \\ 
\hline
 Reflection Index &	$q$ (\r{A}$ ^{-1} $) (\textit{hkl}) & $q$ (\r{A}$ ^{-1} $) (\textit{hkl}) & $q$ (\r{A}$ ^{-1} $) (\textit{hkl}) &	$q$ (\r{A}$ ^{-1} $) (\textit{hkl}) \\ 
 \hline
1	& -	& - & 1.68 (111)  & - \\
2	& -	& - & 1.94 (200)  & - \\
3	& 2.16 (200) & - & - & 2.15 - 2.19 (200) \\
4	& 2.29 (011) & - & - & 2.27 - 2.35 (011)\\
5	& - & 2.69 (111) & - & -  \\
6	& - & 3.10 (200) & - &	-  \\
7	& hidden by Al (200) and InSb (311) & -	& -	& 3.12 - 3.14 (211)  \\
8	& hidden by Al (200) and InSb (311) & -	& -	& 3.15 - 3.17 (220)  \\
9	& -	& - & 3.21 (311) & - \\
10	& 3.95 (002) & - & - & 3.90 - 4.03 (002)  \\
\hline
Lattice parameters	&  & &  &  \\
\hline
a (\r{A}) & 5.82 &	4.04 &	6.48 & 5.73-5.88	 \\
b (\r{A}) & 5.42 &	4.04 &	6.48 & 5.28-5.43	 \\
c (\r{A}) & 3.18 & 4.04 & 6.48 & 3.10-3.22	 \\
\end{tabular}
\end{ruledtabular}
\end{table*}

In order to determine the shape and duration of the strain pulse, the hydrodynamic evolution of the Al film after illumination by the laser pulse was simulated using the code ESTHER \cite{ColombierPRB05}. In this one-dimensional Lagrangian code, the laser energy deposited is calculated by solving the Helmholtz equations, and the evolution of the mass density of the Al film, $\rho_{Al}(z, t)$, is governed by thermodynamics coupled to a multi-phase equation of state. On the time scale of relevance here, only longitudinal expansion takes place, and the strain ($ \sigma $) can be inferred from the mass density: $\sigma(z, t) = \rho_{Al,s}/\rho_{Al}(z, t) - 1$, where $\rho_{Al,s}$ is the density of solid Al, i.e., before the laser pulse. The strain profile in Al at a depth of 300 nm was extracted and an acoustic transmission coefficient from Al to InSb was applied based on the acoustic impedances of the two materials \cite{LoetherSD14,EnquistPRL07}. To reproduce the experimental laser conditions, simulations were performed using a range of absorbed fluences in Al similar to the experimental conditions. The calculated strain in InSb at a depth of about 1 $\mu$m is shown in Figure \ref{fig2}(a). This depth corresponds to a propagation time of 300 ps. The strain shows a very narrow compressive peak (negative strain) that reaches a maximum amplitude of 10\% $ \pm $ 0.5\%, followed by an expansion tail (positive strain) with a small amplitude due to the heating of the Al film. This range of strains was reproduced for the fluences 42 $\pm$ 3.5 mJ/cm$ ^{2} $. The transient compressive peak had a width of 100 nm, which corresponds to a duration of 25 ps given the speed of sound in InSb (3900 m/s \cite{SlutskyPR59}).

The strain generated by a single laser pulse along the [111] direction in InSb was probed experimentally by recording $\theta-2\theta$ diffraction curves before and 300 ps after the interaction of the laser pulse with the Al film (Fig. \ref{fig2}(b)). The central part of the $\theta-2\theta$ curve was not recorded since the high intensity of the (111) reflection could have damaged the detector. The X-ray reflectivity was measured before laser excitation (blue diamonds) to determine the angular response function of the instrument (convoluted $\theta-2\theta$ curve in dot-dashed blue line), which is mainly governed by the monochromator bandwidth ($\Delta E/E$ = 1\%). The X-ray reflectivity was then measured after excitation with the laser pulse (red circles). At angles less than the Bragg angle ($2\theta <$ 12.7$^{\circ}$), the X-ray reflectivity remained unchanged, suggesting that no expansion is associated with the launched strain pulse. However, at angles greater than the Bragg angle ($2\theta <$ 12.7$^{\circ}$), the X-ray reflectivity increased significantly, indicating considerable compression along the [111] direction. The transient nature of the strain was confirmed by recording the reflectivity at $2\theta$ = 13.7$^{\circ}$ at a delay of 1 ns (the black square). The reflectivity decreased by one order of magnitude, which indicates that the strain propagates through the depth probed by the X-rays \cite{SuppMat}.

To corroborate the strain measurements, the strain profile extracted from the hydrodynamic simulation was sent to the Stepanov X-ray server \cite{StepanovSPIE04} to calculate the dynamical $\theta-2\theta$ diffraction curve from InSb at 300 ps (Fig. \ref{fig2}(c)). The calculated dynamical $\theta-2\theta$ curve (dot-dashed red line) was convoluted with a Gaussian function that accounts for the temporal profile of the 100 ps X-ray pulse and the angular response function of the instrument shown in Figure \ref{fig2}(b). Excellent agreement was obtained with the experimental data, which confirms the compression of InSb of up to 10\% $ \pm $ 0.5\%. This compression is obtained in InSb by a pressure ($P$) of 5 GPa, where $P = \sigma \times B_{InSb}$, $\sigma$ being the strain and $B_{InSb} \sim$ 50 GPa the bulk modulus of InSb \cite{LandoltBornstein2001}.

\begin{figure}
\includegraphics{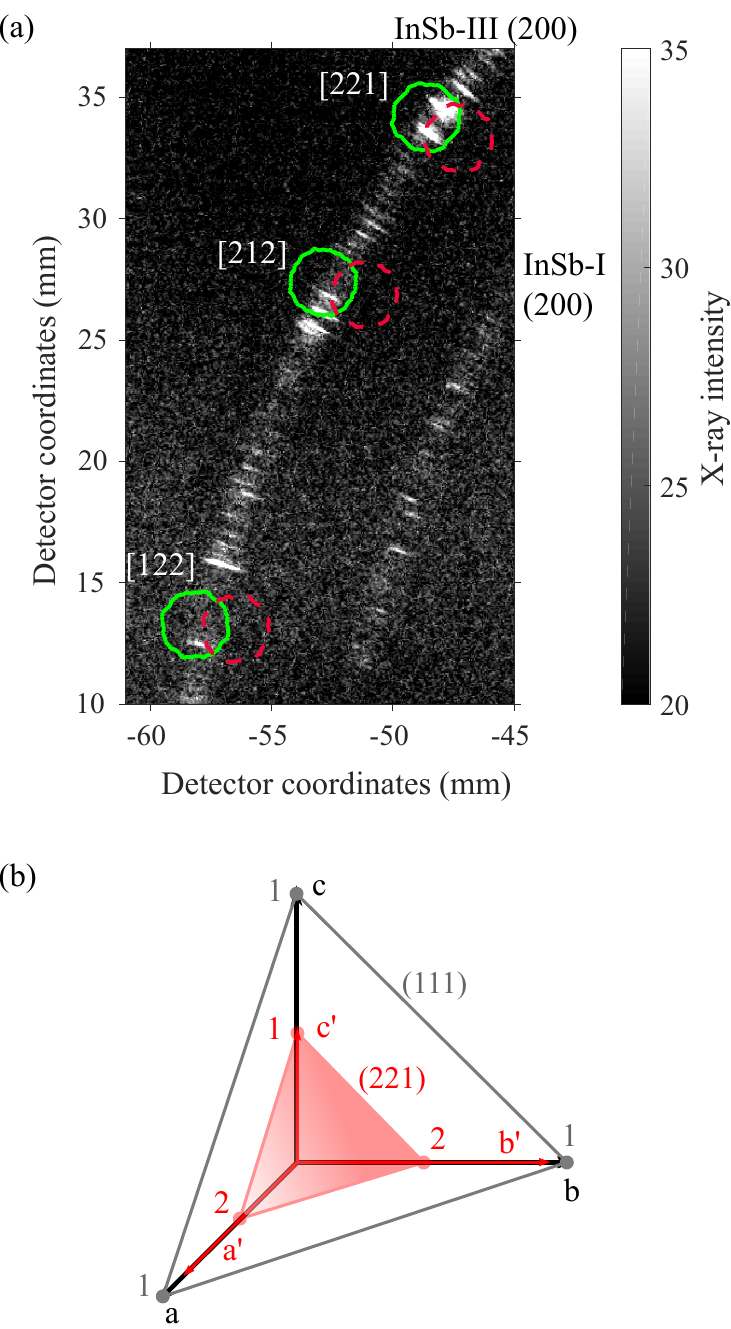}
\caption{\label{fig4} (a) Experimental in situ diffraction pattern and the calculated positions of InSb-III (200) diffraction spots for crystal orientations [122], [212] and [221]. The circles indicate where the reflections would appear for the reported lattice constants. The green circles were calculated using the lowest values of $a^{\prime}$ = 5.73 \r{A}, $b^{\prime}$ = 5.28 \r{A} and $c^{\prime}$ = 3.10 \r{A} \cite{NelmesPRL96}, while the red dashed circles were calculated using the highest values of $a^{\prime}$ = 5.88 \r{A}, $b^{\prime}$ = 5.43 \r{A} and $c^{\prime}$ = 3.22 \r{A} \cite{MezouarPSSB96} (see Table \ref{table1}). (b) Illustration of the (111) plane in the InSb-I cubic cell ($a = b = c$) (gray unshaded triangle) and the (221) plane in the InSb-III orthorhombic unit cell ($a^{\prime} \neq b^{\prime} \neq c^{\prime}$) when $c^{\prime}$ is the short axis (red shaded triangle).}
\end{figure}

After having characterized the pressure pulse generated (5 GPa, 25 ps), \textit{in situ} X-ray diffraction patterns were obtained from the spots on the sample that had been exposed to the laser-generated pressure pulses up to a few days previously. The X-ray angle of incidence was set to 1.3$^{\circ}$ to probe a depth of about 900 nm. At this angle of incidence, the X-ray spot size was 0.85 $ \times $ 0.09 mm$ ^{2} $ (V $ \times $ H). To ensure that the X-ray spot size was smaller than the laser spot, new laser exposures were done leading to permanent damage on an area of 1.4 $ \times $ 0.25 mm$ ^{2} $ (see Supplemental Material \cite{SuppMat}). The \textit{in situ} diffraction pattern was acquired during 10 seconds at an X-ray repetition rate of 1 kHz, which represents 10$^{11}$ incident X-ray photons. Figure \ref{fig3} shows the radial integral of the diffraction pattern as a function of the reciprocal vector \textit{q}, where $ q = 2 \times\pi / d_{hkl} $ and $d_{hkl}$ is the plane spacing of the (\textit{hkl}) reflections. The reflections are indexed with increasing values of \textit{q}, and are identified in Table \ref{table1} for each material. The combination of reflections with indices 3, 4 and 10 can only be found in InSb-III, which confirms that we have indeed stabilized InSb in the InSb-III phase at ambient conditions. From these reflections, the lattice parameters of InSb-III were previously determined for the metastable phase – around 3 GPa – and for the stable phase – above 9 GPa (Table \ref{table1})\cite{NelmesPRB93,NelmesPRL96,MezouarPSSB96}. Despite the large difference in pressure in these previous studies, the lattice parameters were found to be identical. In this study, we obtained the lattice parameters under ambient conditions, and again, our values fall in the same range. To understand why the lattice parameters remain constant over a wide pressure range, further theoretical work is needed on the effect of fast compression/decompression of InSb. 

More information on the formation of the stabilized InSb-III phase can be extracted from the direct \textit{in situ} diffraction pattern on a 2D detector. Part of the diffraction pattern (Fig. \ref{fig4}(a)) shows diffraction rings from the InSb-I (200) and InSb-III (200) planes. The InSb-III (200) ring shows two more intense regions, which are in very good agreement with the calculated positions of the (200) diffraction spots when the InSb-III surface is oriented along [212] and [221]. This indicates that InSb-III is formed along preferred orientations locked to the initial orientation of the InSb-I crystal. Indeed, these preferred orientations correspond to breaking the symmetry of the mono-domain [111] in the InSb-I cubic cell into three equivalent domains, [122], [212] and [221], in the InSb-III orthorhombic cell, when $a$, $b$ or $c$, respectively, is the short axis. Figure \ref{fig4}(b) illustrates that the (111) plane in the InSb-I cubic phase and the (221) plane in the InSb-III orthorhombic phase are nearly parallel when the $c$-axis is the short axis. No diffraction spots were observed from the [122] surface, although in principle, the InSb-I unit cell could transform into the InSb-III unit cell along all three directions ($a$, $b$ or $c$). However, it has been shown experimentally that the application of a uniaxial pressure along the $a$-axis was not able to trigger the transition to the orthorhombic phase \cite{OkaiJPSJ78}. 

In conclusion, we have shown that it is possible to stabilize InSb in the high-pressure orthorhombic phase, InSb-III, using an ultrashort, non-hydrostatic pressure pulse (5 GPa, 25 ps) under ambient conditions. We found that the orthorhombic phase orientations were locked to the initial orientation of the cubic phase. The InSb-III phase was observed to persist for several days. The stability of the high-pressure phase in InSb was investigated theoretically under hydrostatic conditions, however, further theoretical studies on the effect of non-hydrostatic pressure and fast decompression in InSb are required to gain a detailed understanding of the stabilization process of the phase InSb-III. 

\begin{acknowledgments}
We acknowledge financial support from the Swedish Research Council (VR, Grant No. 2014-4732), the Knut and Alice Wallenberg Foundation (KAW, No. 2014.075), Stiftelsen Olle Engkvist Byggmästare (Grant No. 1041513), and the X-PROBE Marie Sk\l{}odowska-Curie ITN within Horizon 2020 (No. 637295). The authors also gratefully acknowledge the assistance of ESRF staff during the beamtime. We would like to thank P. Combris and L. Videau (CEA-DIF, Bruyères-le-Châtel, France) who developed the ESTHER simulation code, M. Harb (Drexel University, PA, USA) who developed the code for calculating the 2D diffraction pattern (available on github: https://github.com/mharb75/Diffraction), and C. Laulhé (Synchrotron SOLEIL, Gif-sur-Yvette, France) for fruitful discussions.
\end{acknowledgments}

\bibliography{InSb_biblio}

\end{document}